\documentclass{kapproc} % Computer Modern font calls

\usepackage{procps} 

\usepackage[dvips]{graphicx}

\upperandlowercase

\kluwerbib % will produce this kind of bibliography entry:

\begin{document}

\articletitle[Galaxy Formation and the Cosmological Angular Momentum Problem]
{Galaxy Formation and the Cosmological Angular Momentum Problem}

\author{Andreas M. Burkert\altaffilmark{1} and Elena D'Onghia\altaffilmark{1,2}} 
 
\affil{\altaffilmark{1}University Observatory Munich, Scheinerstr. 1, D-81679 Munich,
Germany;\\
\altaffilmark{2}Max-Planck-Institut f\"ur Extraterrestrische Physik, Karl-
Schwarzschild-Str. 1, D-85741 Garching, Germany}

\begin{abstract}
The importance of angular momentum in regulating the sizes of galactic disks and by this
their star formation history is highlighted. Tidal torques and accretion of satellites
in principle provide enough angular momentum to form disks with sizes that are in
agreement with observations. However three major problems have been identified
that challenge cold dark matter theory and affect models of galaxy evolution:
(1) too much angular momentum is transferred from the gas to the dark halos during infall, 
leading to disks with scale lengths that are too small, 
(2) bulgeless disks require more specific angular momentum than is generated cosmologically
even if gas would not lose angular momentum during infall,
(3) gravitational torques and hierarchical merging produce a specific angular 
momentum distribution that does not match the distribution required to form exponential 
disks naturally; some gas has exceptionally high angular momentum, leading to extended outer
disks while another large gas fraction will contain very little
specific angular momentum and is expected to fall into the galactic center, forming 
a massive and dominant bulge component.
Any self-consistent theory of galaxy formation will require
to provide solutions to these questions. Selective mass loss of low-angular-momentum
gas in an early phase of galaxy evolution currently seems to be the most promising
scenario. Such a process would have a strong affect on the early protogalactic evolution phase,
the origin and evolution of galactic morphologies and link central properties of
galaxies like the origin of central massive black holes with their global structure.
\end{abstract}

\begin{keywords}
galaxies: disks, formation, kinematics and dynamics, structure - dark matter
\end{keywords}

\section{Introduction}

The origin of the distribution of mass and angular momentum in disk galaxies is yet an 
unsolved astrophysical puzzle. Eggen, Lynden-Bell and Sandage (1962) argued that spiral galaxies like
the Milky Way formed by rapid infall of an initially uniform sphere of gas into a 
centrifugally supported disk. It was soon realized that these disks would have 
characteristic exponential surface density distributions if the initial gas sphere would
be in solid body rotation and if the gas would preserve its initial specific angular momentum
distribution $M(<j)$ during infall (Mestel 1963, Crampin \& Hoyle 1964, Innanen 1966, Freeman 1970).
Here, $M(<j)$ is the cumulative mass of gas with angular momentum less or equal to $j$.

Since these first pioneering studies our insight into galaxy formation has substantially 
changed and improved. Current cosmological models consider a dissipationless cold dark matter (CDM) 
component to dominate structure formation in the Universe (Blumenthal et al. 1984). Initially
small dark matter density perturbations in the early Universe decouple from the Hubble flow,
collapse into virialized dark matter halos and merge into larger and larger structures.
Gas accumulates within the extended dark halos, dissipates its kinetic energy and settles
into the equatorial plane as soon as centrifugal equilibrium is being reached,
forming fast rotating disks that subsequently turn into stars (White \& Rees 1978).
Fall \& Efstathiou (1980) argued that gas and dark matter should initially be well mixed. In this case,  
the specific angular momentum distribution of the gas should be
equal to that of the dark halo. If $M(<j)$ would be preserved during infall into the equatorial
plane, the exponential disk scale length $R_d$ would be directly related to the 
specific angular momentum $\lambda$ of the dark halo. Here
$\lambda$ is the dimensionless spin parameter (e.g. Peebles 1969)

\begin{equation}
\lambda = \frac{J |E|^{1/2}}{G M_{vir}^{5/2}}
\end{equation}

\noindent where J, E, and $M_{vir}$ are
the total angular momentum, energy and virial mass of the halo, respectively and G is
Newton's constant.  This shifted the focus to a more detailed investigation of
the spin properties of dark halos and a determination of $\lambda$. 
Peebles (1969) suggested that halos would acquire non-negligible specific angular momentum by the
gravitational interaction of their building blocks with neighboring structures.
Subsequent cosmological
N-body simulations (Barnes \& Efstathiou 1987, Efstathiou \& Barnes 1983, Zeldovich \& Novikov 1983,
Cole \& Lacey 1996) confirmed this mechanism. They showed that prior to collapse the 
angular momentum of a dark fluctuation grows roughly linearly with time,
as predicted by linear tidal-torque theory (White 1984) until it
decouples from the Hubble flow, collapses and virializes. The $\lambda$ distribution of virialized
halos turned out to be well described by a log-normal (Steinmetz \& Bartelmann 1995, Cole \& Lacey
1996, Gardner 2001, Bullock et al. 2001)

\begin{equation}
p( \lambda ) d \lambda = \frac{1}{\sigma_{\lambda} \sqrt{2 \pi}} \exp \left(-\frac{
ln^2(\lambda/\lambda_0)} {2 \sigma^2_{\lambda}} \right) d ln \lambda
\end{equation}

\noindent with median value of $\lambda_0=0.042 \pm 0.006$ and dispersion 
$\sigma_{\lambda}=0.5 \pm 0.04$. A more partical spin parameter $\lambda'$ was
proposed by Bullock et al. (2001):

\begin{equation}
\lambda'=\frac{J}{\sqrt{2}M_{vir}V_{vir}R_{vir}}
\end{equation}

\noindent with $R_{vir}$ the halo virial radius and $V^2_{vir}=GM_{vir}/R_{vir}$  its
virial velocity. The spin parameters
$\lambda$ and $\lambda'$ turn out to be very similar due to the fact that dark halos are well
described by a  universal density distribution (Navarro, Frenk \& White, 1997, NFW) and the
$\lambda'$ distribution also follows a log-normal with best fit values $\lambda'_0=0.035 \pm 0.005$
and $\sigma_{\lambda'}=0.5 \pm 0.03$ (Bullock et al. 2001). 

Given $\lambda'$, and assuming that the disk gas will have the same specific angular momentum
as the dark halo, the exponential disk scale length $R_d$ can be easily determined. Adopting a flat
rotation curve with velocity $v_c$, the disk's specific angular momentum is (Mo et al. 1998)

\begin{equation}
j_d = 2 R_d v_c = \sqrt{2} \lambda' V_{vir} R_{vir} .
\end{equation}

\noindent For typical NFW halos with  concentrations $c=10$, the peak velocity 
of the dark matter rotation curve which is in general of order
the observed peak rotation velocity is $v_{peak} = 1.2 V_{vir} \approx v_c$. For a flat
$\Lambda$CDM-cosmology with cosmological parameters $\Omega_m=0.3$, $\Omega_{\Lambda}=0.7$ 
and a Hubble constant of 70 km s$^{-1}$ Mpc$^{-1}$ the virial parameters
of dark halos are coupled by the relations (Navarro \& Steinmetz 2000)

\begin{equation}
R_{vir} = 270 \times \left(\frac{M_{vir}}{10^{12}M_{\odot}} \right)^{1/3} kpc  = 
215 \times \left(\frac{V_{vir}}{100 km/s} \right) kpc .
\end{equation}

\noindent Combining these questions leads 
to a relationship between the exponential disk scale length $R_d$ and the disk's
rotational velocity $v_c$:

\begin{equation}
R_d = 8.5 \left( \frac{\lambda'}{0.035} \right) \left(\frac{v_c}{200 km/s} \right) kpc .
\end{equation}

\noindent A more detailed investigation which takes into account adiabatic contraction of the dark
halo (Jesseit et al. 2002) and the combined gravitational force of the disk and dark halo
shows that equation 6 overestimates the disk scale length by a small amount of order 
10\% to 20\%.

\begin{figure}[ht]
\vskip.2in
\centerline{\includegraphics[width=4in]{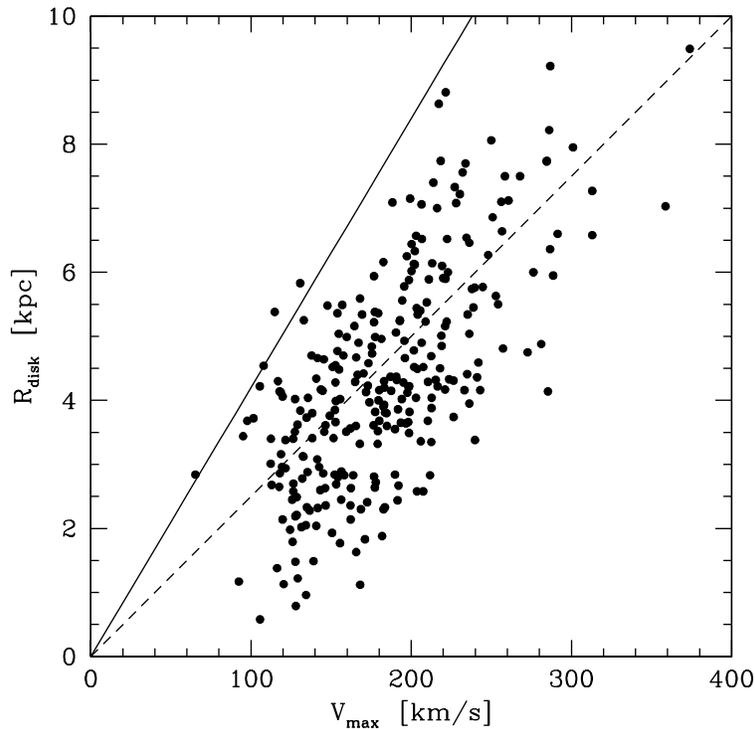}}
\caption{The observed disk scale lengths versus their maximum rotation velocities are shown for
the Courteau (1997) sample. The solid line shows the theoretically predicted correlation
for $\lambda' = 0.035$. The dashed curve corresponds to $\lambda' = 0.025$}
\end{figure}

Figure 1 compares the observations (Courteau 1997) 
with the prediction of equation 6, adopting a typical
value of $\lambda'=0.035$. Note that the slope reflects directly the spin parameter $\lambda'$.
The theoretical predictions lead to a correlation that is steeper than observed. 
The dashed line shows a best fit model which requires a value of
$\lambda'=0.025$. If the rotation speeds of galactic disks are approximately the same as 
the maximum circular velocities of their dark halos, and if the gas would have the same
specific angular momentum as the dark halo, cosmological models predict
disk scale lengths that are about a factor of 1.4 
larger than observed. Good agreement could be
achieved if the gas retained only 70\% of the available angular momentum during
infall. A similar conclusion was reached by Navarro \& Steinmetz  (2000) and Mo, Mao \& White (1998) who
investigated the structural properties of galactic disks within characteristic
NFW-type dark halos in greater details.

\section{Angular Momentum Loss during Galactic Disk Formation}

Numerical simulations have become one of the most powerful tools for
exploring galaxy formation. Hydrodynamical simulations of disc 
formation were initiated by Navarro $\&$ Benz (1991), who included 
radiative
cooling by hydrogen and helium, and attempted to account for star 
formation
and feedback processes. In contrast to the conclusion drawn in the previous chapter,
the simulated galaxies however failed to
reproduce their observed counterparts: the disks were
found to be too small and were more centrally concentrated than actual
galaxies (Navarro $\&$ White 1993). In addition, star formation in the
models was overly efficient, converting gas into stars too fast.
Many of the shortcomings of the early modeling could be accounted for by
the limited resolution of the simulations and the way
in which feedback was treated.  As star formation 
was very efficient in low mass halos at high redshifts, 
a large number of dense, compact stellar systems formed that lateron
were collected in the innermost regions of larger galaxies that formed
through the merging of smaller objects. 
Angular momentum loss by dynamical friction drove these star clusters to the center
where they formed large, dense bulges instead of extended disks.

\begin{figure}[ht]
\vskip.2in
\centerline{\includegraphics[width=4in]{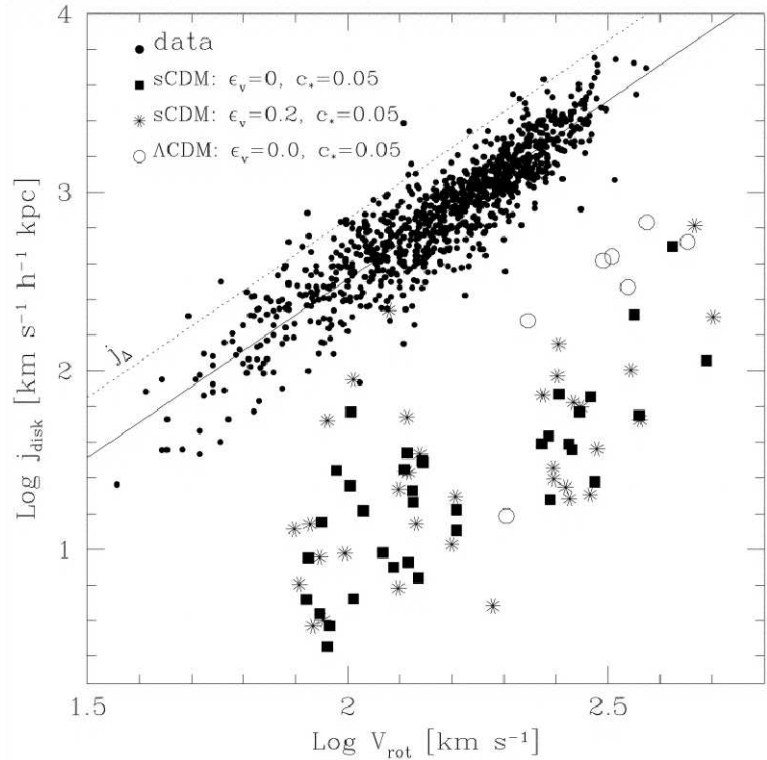}}
\caption{The specific angular momentum $j_{disk}=R_d \times v_{rot}$ of model disks
with scale length $R_d$ and rotational velocity $v_{rot}$ is compared with
observational data (filled points). Figure adapted from Navarro \& Steinmetz (2000).}
\end{figure}

The cosmological angular momentum problem was reinvestigated lateron in 
greater details by Navarro $\&$ Steinmetz (2000) with high-resolution N-body/gas\-dynamical
simulations that included star formation and feedback to examine the 
origin of the I-band Tully-Fisher relation for different comologies. Although
the slope and the scatter could be well reproduced in the simulations, the models
failed to match the zero point (Fig. 2). Again, the galaxies were too compact with respect
to the observations even with realistic feedback formulations that were calibrated
to reproduce the empirical correlations, connecting the local star formation rate with 
the gas surface density (Kennicutt 1998).

A possible solution is suppression of early cooling of the gas by strong feedback from supernovae
which prevents drastic angular momentum loss and produces better
fits to the observations (Sommer-Larsen et al. 2003, Abadi et al. 2003).
However, even in these simulations the disk systems typically contained
denser and more massive bulges than observed late-type galaxies, indicating that
the specific angular momentum problem is not completely solved.

\section{Spin Parameter and Halo Merging History}

In the Fall $\&$ Efstathiou (1980) model, the scale size of a galaxy is 
determined by its angular momentum, which is acquired by tidal torques from 
neighboring objects in the expanding universe, prior to the collapse of the halo. 
Recent results from numerical N-body simulations
have suggested that major mergers might be especially important to 
increase the mean angular momentum content of the halos and their spin 
parameters (Gardner 2001; Vitvitska et al. 2002). This is due to the
substantial amount of orbital angular momentum which a major merger adds
to the system and which dominates
the final net angular momentum of the remnant (Gardner 2001).
Minor mergers, in contrast, have little effects on the angular momentum budget of
galaxies.

Semi-analytical models of galaxy evolution (Kauffmann et al. 1993) cannot investigate 
the angular momentum properties of a halo self-consistently.  They can however 
predict their merging history and by this their spin parameters, if the spin
is coupled with the history of minor and major mergers. This question has
been investigated in details by Vitvitska et al. (2002). 
They followed the evolution of the cumulative mass and the spin 
parameter of the major progenitors for three high-resolution halos and found
that the spin parameters clearly change with time. Instead of a gradual increase,
$\lambda$ is increasing with every major merger and subsequently decreasing as a
result of minor merging. They proposed that the net
angular momentum of dark matter halos originates from statistically
random merging with preferentially the last major merger dominating.
In this picture, the evolution of the halo angular momentum is quite 
different
from the standard tidal torques scenario, in which the angular momentum 
grows
steadily at early times and its growth flattens lateron.

\begin{figure}[ht]
\vskip.2in
\centerline{\includegraphics[width=4in]{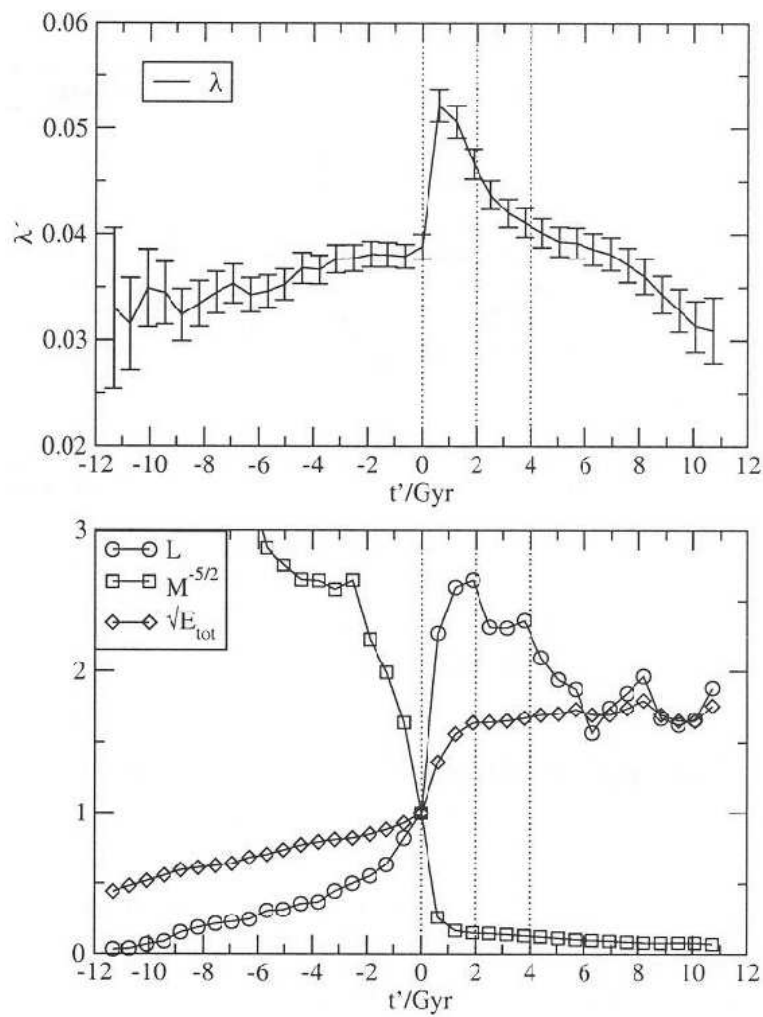}}
\caption{The dimensionless spin parameter $\lambda$ is shown in the upper panel for
dark halos which experienced a  major merger. The lower panel shows the evolution 
of the total angular momentum $L$, the total mass $M$ and the total energy $E_{tot}$ with time.
The time axis $t'$ is centered on the epoch of major merging.}
\end{figure}

This conclusion has recently been questioned by Hetznecker \& Burkert (in preparation). Figure 3
shows the evolution of the mean spin parameter $\lambda'$, averaged over a large set of
dark halos that experienced at least one major
merger in their lifetime. The time axis is centered at the epoch of the major merging.
Note that indeed a substantial increase in $\lambda'$ is visible when the merger occurs.
However, afterwards, within the next 2 Gyrs, $\lambda'$ decreases again, approaching
the same low value that had been achieved prior to the merging event. Hetznecker \& Burkert
argue that the temporary increase in $\lambda'$ is a result of the fact that dark halos
are out of virial equilibrium during this epoch. 
Dark particles oscillate into and out of
the virial radius and the most bound particle, with respect to which the angular momentum is
typically defined, is not a good measure of the center of mass.
Neglecting unrelaxed dark halos, the authors find no difference between major and minor mergers
and an average $\lambda'$-distribution that is on average shifted to somewhat lower values
compared with previous studies that include unrelaxed halos.

\section{A Test Case: Bulgeless Galaxies} 

Most work has up to now focused on angular momentum properties of
halos that had at least one dominant major merger during their evolution.
However, major merger events tend to destroy disks, producing
spheroidal stellar systems, like bulges or early-type galaxies (see e.g.  
review by
Burkert \& Naab 2003). Not much work has been devoted to explore the 
angular momentum
properties of halos that host pure disk galaxies, or bulgeless
galaxies and that never experienced a major merger.

Recently D'Onghia $\&$ Burkert (2004) performed three N-body simulations 
in a $\Lambda$CDM cosmological universe and explored the angular momentum 
properties
of halos that did not experience any major merger from
redshift 3 until the present time and that are in principle good 
candidates to
host bulgeless galaxies. The authors traced each identified halo 
backward in time, following the mass
of the most massive progenitor as a function of redshift during $0<z<3$ 
(for details, see D'Onghia $\&$ Burkert, these proceedings). They
found that disk-dominated late-type galaxies inhabiting
halos that have not experienced a major merger have a
distribution of $\lambda'$ that peaks around
a value of 0.023  which is substantially smaller than expected from 
observed rotation curves of bulgeless disks. 
 To demonstrate this they compared their results with
the sample of
van den Bosch, Burkert $\&$ Swaters (2001) who determined spin
parameters for 14 late-type bulgeless disk galaxies.

\begin{figure}[ht]
\vskip.2in
\centerline{\includegraphics[width=4in]{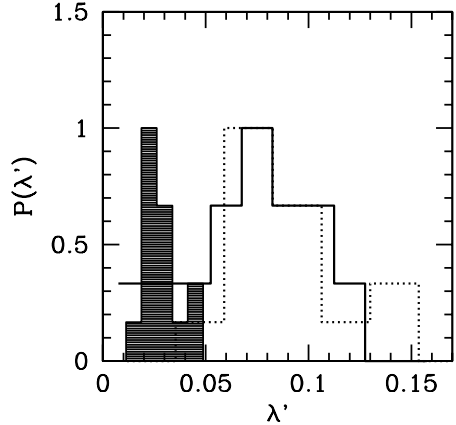}}
\caption{The distribution of the spin parameters $\lambda'$ of dark halos that experienced
no major merger is shown by the filled histogram and compared to the spin distribution 
of observed bulgeless disk galaxies (solid line). 
The dotted curve shows the $\lambda'$ distribution if the spins of
dark halos without major mergers would be multiplied by a factor of 3.15.}
\end{figure}

Fig.4 shows the probability distribution of the spin
parameter of halos that did not experience any major mergers since z=3
(filled region) and compares it with the normalized probability 
distribution
of $\lambda'_{disk}$ for the sample of galaxies measured by van den Bosch, 
Burkert $\&$ 
Swaters (2001) assuming a mass-to-light ratio of unity in the R band.
The galaxies show a distribution that follows a lognormal 
distribution with
an average value for $\lambda'_{disk} \approx 0.067$, and a dispersion of
$\sigma_{\lambda'}\approx 0.31$. This is a factor of 3 larger than 
predicted
by the numerical simulations. Again, galactic disks forming in these halos
would be too small. However,
this time, the problem cannot be solved by feedback processes which 
prevent
the loss of specific angular momentum of clumpy infalling gas.
It is remarkable that multiplying the $\lambda'$ values of simulated
halos by a factor 3.15 (dotted histogram in Fig. 4) reproduces the peak
and dispersion of the observed $\lambda'_{disk}$ distribution quite well.
Halos without major mergers
acquire their specific angular momentum through tidal torques
in the early epochs of evolution (Barnes $\&$ Efsthatiou 1987),
when the density contrasts were small, in
accordance with the prediction of the linear theory.
The net result are halos with typical spin parameters of $\lambda' = 0.02$
whereas data for bulgeless galaxies indicate halos with values of 
$\lambda' = 0.06 - 0.07$,
pointing out a new angular momentum problem, especially for bulgeless galaxies.

It's not clear how to overcome this problem. 
It is known that the spin parameter distribution for the
collapsed objects is insensitive to the shape of the initial power
spectrum of density fluctuations, to the environment
and the adopted cosmological model (Lemson $\&$ Kauffmann 1999).

In the Fall $\&$ Efsthatiou (1980) model it is assumed that the gas and the dark 
matter in
a protogalaxy have the same distribution of specific angular momentum.
However the two components undergo different relaxation mechanisms: the 
dark matter experiences collisionless violent relaxation and the gas shocks
and dissipates its kinetic energy and settles into the central regions. 
Van den Bosch et al. (2002)
used numerical simulations in a cold dark matter cosmology to compare the 
angular momentum distributions of dark matter and non-radiative gas.
They showed that gas and 
dark matter have identical angular momentum distributions after the
protogalactic collapse phase if, and only if, cooling is ignored, in agreement 
with the standard assumptions. In addition, they found that 
5\% to 50\% of the mass has negative specific angular momentum. 
Realistic disks do not typically contain counterrotating material.
This suggests that during the cooling process the gas with negative 
specific angular momentum collides with material with positive specific 
angular momentum to build a bulge component. Thus, even without 
substructures
and consequent dynamical friction a bulge would be likely to form in all galaxies
due to this process, making the formation of bulgeless 
galaxies even more difficult.

\section{The Angular Momentum Problem of Low-Mass Elliptical Galaxies}

\begin{figure}[ht]
\vskip.2in
\centerline{\includegraphics[width=4in]{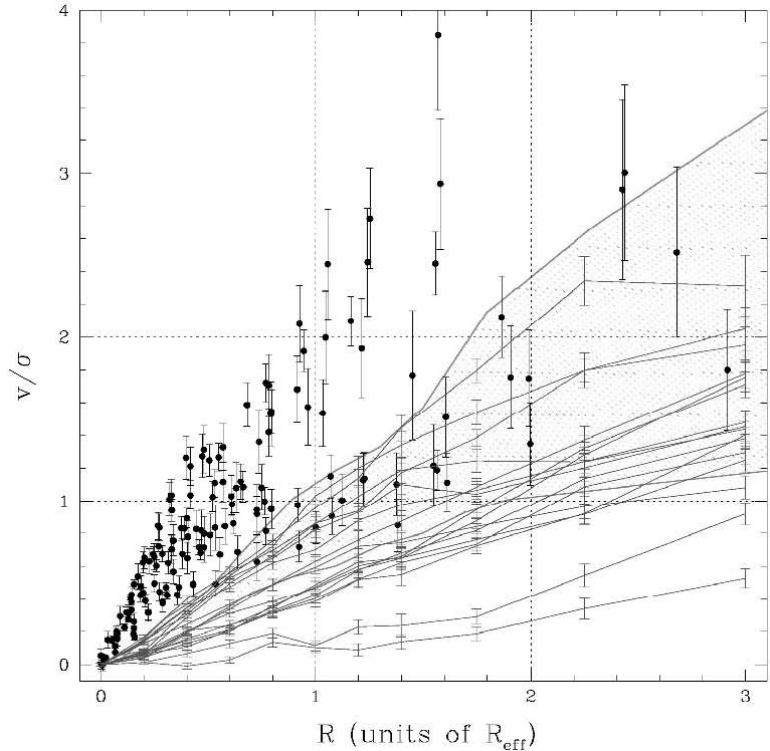}}
\caption{Comparison of observed $v/\sigma$ (circles with error bars) and $v/\sigma$
for edge-on numerical merger remnants resulting from collisions of disk galaxies.
The shaded area corresponds to the range occupied by Bendo \& Barnes (2000).
Both distributions are not compatible with the observed one, even in this edge-on case.
}
\end{figure}

As outlined previously, numerical simulations indicate an angular momentum problem 
for disk galaxy formation that results from the fact that gas loses
a substantial fraction of its angular momentum while settling into
the equatorial plane. Cretton et al. (2001) however also found a related problem for
fast rotating low-mass elliptical galaxies. Figure 5 shows the observed radial dependence
of $v/\sigma$ of their early-type galaxy sample with $v$ the observed 
line-of-sight rotational velocity of the stellar spheroid along
the major axis and $\sigma$ the local line-of-sight
velocity dispersion. The shaded region and the solid lines show the results
of numerical simulations of unequal and equal-mass spiral galaxy mergers which reproduce
many of the global properties of ellipticals remarkably well (Naab \& Burkert 2003, 
Burkert \& Naab 2003). Note that none of these models leads to curves
that rise as steeply as observed for many cases of low-mass ellipticals which
show values of $v/\sigma \approx 2$ within
1-2 effective radii. Theoretical models instead predict values of $v/\sigma \approx 1$.

Like in the case of galactic disk formation, tidal torques and dynamical friction
during the major merger event efficiently remove specific angular momentum from the baryonic stellar
component, leading at the end to stellar spheroids with rotational velocities that do not exceed 
much the velocity dispersion.

\section{The Specific Angular Momentum Distribution Problem}

\begin{figure}[ht]
\vskip.2in
\centerline{\includegraphics[width=4in]{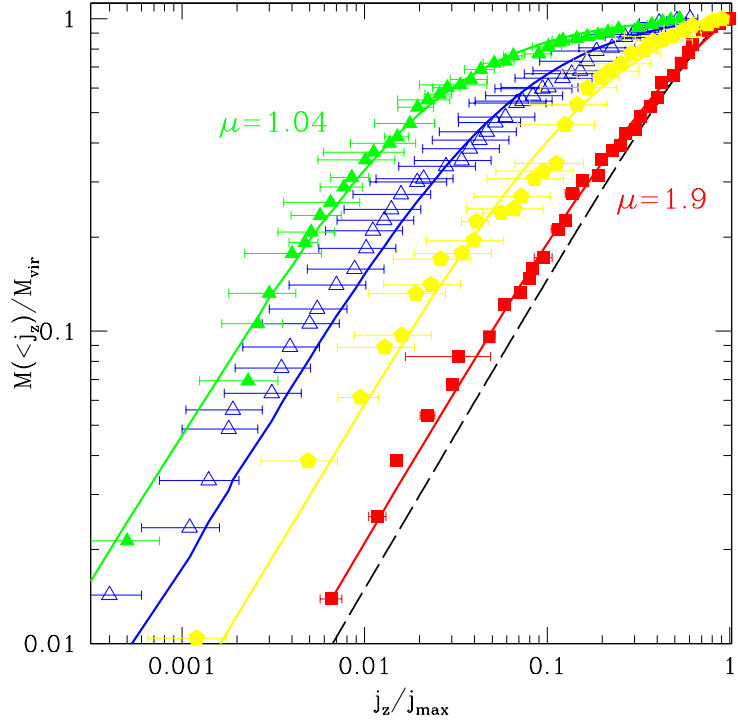}}
\caption{The total mass $M(<j)$ of dark matter with angular momentum less than $j$
is fitted by equation (7). Figure adapted from Bullock et al. (2001).}
\end{figure}

Early estimates assumed that the detailed specific angular momentum distribution of dark halos
is well described by a hypothetical uniform sphere in solid body rotation. This question was
reinvestigated again by Bullock et al. (2001) using a large statistical sample of halos that was
drawn from a high-resolution $\Lambda$CDM simulation. They found that the angular 
momentum distribution of dark
halos has indeed a universal form which however strongly deviates from the previously expected
uniformly rotating sphere. The angular momentum distribution can be well fitted by (Fig. 6)

\begin{equation}
M(<j)=M_{vir} \frac{\mu j}{j_0+j}
\end{equation}

\noindent where $M(<j)$ is the cumulative
total mass of dark matter with angular momentum less than
$j$ and $\mu > 1$ is a free shape parameter. The
characteristic specific angular momentum $j_0$ is determined by

\begin{equation}
j_0 = \sqrt{2} V_{vir} R_{vir} \lambda' / b(\mu)
\end{equation}

\noindent with

\begin{equation}
b(\mu) = -\mu \ln (1-\mu^{-1})-1.
\end{equation}

Note, that this formula ignores any material with negative angular momentum 
(Chen et al. 2003) which makes the problem of low-angular momentum gas even 
worse, as discussed in chapter 5.

Following Bullock et al. (2001) we explore the surface density distribution
of galactic disks, adopting $\lambda' = 0.04$, a NFW halo with concentration $c=14$ and 
a disk baryon fraction of $f_{disk}=M_{disk}/M_{vir}=0.03$. We keep 
$\mu$ as a free parameter within the measured range of $1.01 \leq \mu \leq 2$ and
study its effect on the structure of the disk.
Assuming that
the dark halo contracts adiabatically due to the gravitational force of the
infalling gas, one can calculate the radius $r_*$ in the equatorial plane where gas of a
given angular momentum $j_*$ reaches centrifugal equilibrium. $r_*$ is given
by the implicit equation

\begin{equation}
j_*=Gr_*(M_{NFW}(r_{cont})+M_{gas}(<j_*))
\end{equation}

\noindent where $r_{cont}$ is the initial radius of the dark halo mass shell
that after adiabatic contraction ends up at $r_*$
(Jesseit et al. 2002; Blumenthal et al. 1984):

\begin{equation}
r_{cont}=r_* \frac{M_{NFW}(r_{cont})+M_{gas}(<j_*)}{M_{NFW}(r_{cont})}.
\end{equation}

\begin{figure}[ht]
\vskip.2in
\centerline{\includegraphics[width=4in]{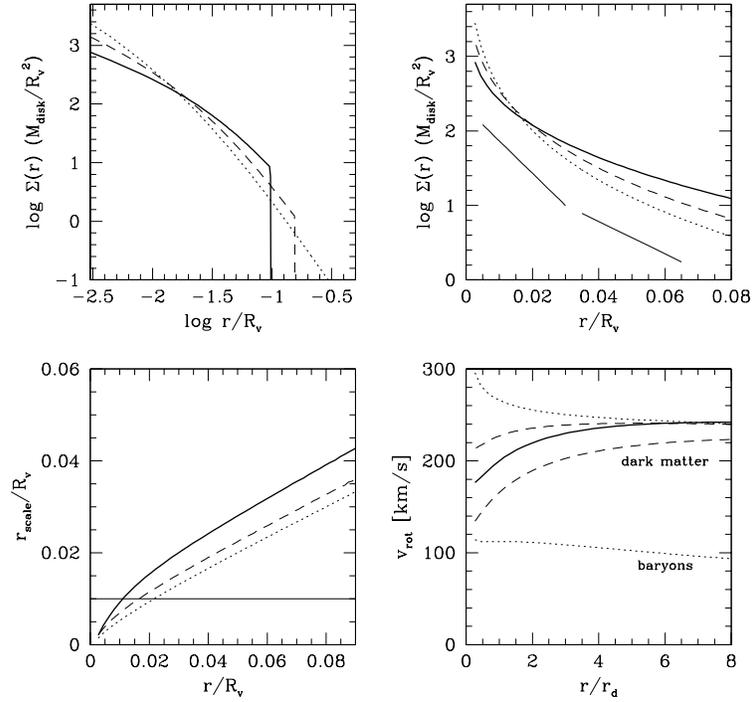}}
\caption{The predicted disk surface density distribution is shown for disks that
form from gas with a specific angular momentum profile as predicted by equation 7. The dotted,
dashed and solid lines show cases with $\mu=1.06$, $\mu=1.25$ and $\mu=2.0$, respectively.
The upper left panel shows a log-log representation. The upper right panel shows
the corresponding log-linear plot and compares it with an exponential disk of scale length
$0.01 R_{v}$ and $0.02 R_{v}$, where $R_v$ is the dark halo virial radius. 
The lower left panel shows the local disk scale length, normalized to the virial radius.
Typical observed scale lengths are $0.01 R_v$. The lower right panel shows the corresponding
rotation curves adopting a dark halo mass of $10^{12} M_{\odot}$, which is characteristic
for the Milky Way.}
\end{figure}

The upper left panel of figure 7 shows the normalized disk surface density profiles for
$\mu = 1.06$, $\mu = 1.25$ and $\mu = 2.0$. The profiles agree with Bullock et al. (2001,
see their Fig. 20). In this representation it is difficult to estimate which profile,
if any, would fit an exponential disk. Therefore  the upper right panel 
shows again the same profiles, however now in a log-linear representation
and focussing on the observed exponential disk regime of $r/R_d \leq 4$, 
where a disk scale length of $R_d \approx 0.01-0.02 R_{vir}$
has been assumed. An exponential disk with scale length
of $0.01 R_{vir}$ or $0.02 R_{vir}$ is shown by the straight thin solid lines.
Even for large values of $\mu \geq 2$ the profiles are not well fitted
by an exponential. All profiles instead rise above the exponential in the
inner and outer regions. The situation becomes
even more clear in the lower left planel which shows the local exponential scale length
defined as $r_{scale} = (dln\Sigma/dr)^{-1}$ for all three values of $\mu$.
In all cases, $r_{scale}$ is continuously and steeply increasing with radius,
with no sign of a plateau at a characteristic scale length of order $0.01 - 0.02 R_{vir}$.
If one assumes that all material with small scale length $r_{scale} < 0.01 R_{vir}$
forms a bulge component one
can calculate the predicted bulge-to-disk ratio as function of $\mu$. We find that for all 
reasonable values of $\mu$, galactic disks should harbour large bulges with 
bulge-to-disk mass ratios of more than 20\% which is not consistent with the 
population of late-type disk galaxies.

Viscous effects and secular evolution might change the surface density profiles
of galactic disks and redistribute their specific angular momentum distribution.
It has indeed been shown e.g. by Slyz et al.  (2002) that exponential stellar disks
would arise naturally from rather arbitrary initial conditions if the star formation timescale 
is equal to the viscous timescale. It is however unlikely that viscous effects would
increase the specific angular momentum of the gas in the innermost regions, by this reducing
the bulge mass fraction.

\section{Summary}
Although, on average, cosmological models of structure formation in a $\Lambda$CDM universe
generate enough spin to explain the origin of extended galactic disks several problems
still remain to be solved. These include the loss of  angular momentum during gas infall,
the specific angular momentum distribution and exponential disk formation and the origin 
of bulge-less galaxies. The last problem might be even more puzzling if many bulges formed
by secular processes as summarized recently in Kormendy \& Kennicutt (2004). 
Feedback has been invoked as a mechanism to prevent
the process of drastic angular momentum loss of infalling gas.
(van den Bosch, Burkert $\&$ Swaters 2002; 
Maller $\&$ Dekel 2002; Maller, Dekel $\&$ Somerville 2002).
However, D'Onghia $\&$ Burkert (2004)  have shown that the dark
halos that experienced no major mergers have already too low an angular 
momentum to
produce the observed disks and it is not clear which feedback process  
would increase
the specific angular momentum of the gas beyond that of the dark component
in order to explain this result.
Selective outflow of especially low-angular momentum
gas could provide another solution. Again it seems difficult to understand bulgeless galaxies
in this context as in general a bulge component would be required to generate the kinetic
energy to drive galactic winds. Another interesting scenario was proposed by
Katz et al. (2003) and Birnboim and Dekel (2003) who suggested that
gas in high-redshift disk galaxies is accreted cold and without virial shocks 
directly from filaments. 
More work along these lines is  clearly required to understand the origin of galaxies
and their angular momentum  in greater details.

\begin{acknowledgments}
The authors wish to thank David Block and Ken Freeman for organising a conference on Penetrating
Bars through Masks of Cosmic Dust.
\end{acknowledgments}


\begin{chapthebibliography}{1}
\bibitem{}
Abadi, M.G., Navarro, J.F., Steinmetz, M., Eke, V.R. 2003, ApJ, 591, 499 
\bibitem{}
Barnes, J.E., $\&$ Efsthatiou, G. 1987,  ApJ, 319, 575
\bibitem{}
Bendo, G. \& Barnes, J. 2000, MNRAS, 316, 315
\bibitem{}
Birnboim, Y., Dekel, A. 2003, MNRAS, 345, 349
\bibitem{}
Blumenthal, G.R., Faber, S.M., Primack, J.R. \& Rees, M.J. 1984, Nature, 311, 527
\bibitem{}
Bullock, J.S., Dekel, A., Kolatt, T.S.,  Kravtsov, A.V., Klypin, A.A., 
 Porciani, C., $\&$ Primack, J.R. 2001, ApJ, 555, 240
\bibitem{}
Burkert, A. $\&$ Naab, T. 2003, in Galaxies and Chaos, eds. G. Contopoulos 
\& N. Voglis, Springer, p. 327
\bibitem{}
Chen, D. N., Jing, Y.P., Yoshikaw, K.  2003, ApJ, 597, 35
\bibitem{}
Cole, S. \& Lacey, C. 1996, MNRAS, 281, 716
\bibitem{}
Courteau, S. 1997, AJ, 114, 2402
\bibitem{}
Crampin, D.J, \& Hoyle, F. 1964, ApJ, 140, 99
\bibitem{}
Cretton, N., Naab, T., Rix, H.-W. \& Burkert, A. 2001, ApJ, 554, 291
\bibitem{}
de Jong, R.S., Lacey, C. 2000, ApJ, 545, 781
\bibitem{}
D'Onghia, E. $\&$ Burkert, A. 2004, ApJL submitted, (astro-ph/0402504)
\bibitem{}
D'Onghia, E. $\&$ Burkert, A. 2004, this procceding.
\bibitem{}
Efstathiou, G. \& Barnes, J. 1983, in Proc. 3rd Moriond Astrophys. Meeting, 
Formation and Evolution of Galaxies and Large Structures in the Universe, ed.
J. Audouze \& J. Tran Thanh Van (Dordrecht: Reidel), p361
\bibitem{}
Eggen, O.J., Lynden-Bell, D. \& Sandage, A.R. 1962, ApJ, 136, 748
\bibitem{}
Fall, S.M., Efsthatiou, G. 1980, MNRAS, 193, 189 
\bibitem{}
Freeman, K.C. 1970, ApJ, 160, 811
\bibitem{}
Gardner, J.P. 2001, ApJ, 557, 616
\bibitem{}
Jesseit, R., Naab, T. \& Burkert, A. 2002, ApJ, 571, L89
\bibitem{}
Innanen, K.A. 1966, AJ, 71, 64
\bibitem{}
Katz, N., Keres, D., Dave, R., Weinberg, D.H. 2003, in "The IGM/Galaxy 
Connection: The Distribution of Baryons at z=0", Vol. 281., eds. 
Jessica L. Rosenberg and Mary E. Putman
\bibitem{}
Kauffmann, G., White, S.D.M. \& Guiderdoni, B. 1993, MNRAS, 264, 201
\bibitem{}
Kennicutt, R.C. 1998, ARAA, 36, 189
\bibitem{}
Kormendy, J. \& Kennicutt, R.C. 2004, ARAA, in press
\bibitem{}
Lemson, G., $\&$ Kauffmann, G. 1999, MNRAS, 302, 111
\bibitem{}
Maller, A.H., $\&$ Dekel, A. 2002, MNRAS, 335, 487
\bibitem{}
Maller, A.H., $\&$ Dekel, A., Somerville, R. 2002, MNRAS, 329, 423
\bibitem{}
Mestel, L. 1963, MNRAS, 126, 553
\bibitem{}
Mo, H.J., Mao, S. \& White, S.D.M. 1998, MNRAS, 295, 319
\bibitem{}
Naab, T., \& Burkert, A. 2003, ApJ, 597, 893
\bibitem{}
Navarro, J. $\&$ Benz, W.  1991, ApJ, 380, 320
\bibitem{}
Navarro, J., Frenk, C.S. \& White, S.D.M. 1997, ApJ, 490, 493
\bibitem{}
Navarro, J.F., Steinmetz, M. 2000, ApJ, 538, 477
\bibitem{}
Navarro, J. $\&$ White, S.D.M. 1993, MNRAS, 265, 271
\bibitem{}
Peebles, P.J.E. 1969, ApJ, 155, 393
\bibitem{}
Slyz, A.D., Devriendt, J.E.G-, Silk, J. \& Burkert, A. 2002, MNRAS, 333, 894
\bibitem{}
Sommer-Larsen, J., G\"otz, M., Portinari, L. 2003, ApJ, 596, 47
\bibitem{}
Steinmetz, R.S. \& Bartelmann, M. 1995, MNRAS, 272, 570
\bibitem{}
Van den Bosch, F.C., Burkert, A., $\&$  Swaters, R.A. 2001, MNRAS, 326, 
1205
\bibitem{}
Van den Bosch, F., Abel, T., Croft, R.A.C., Hernquist, L. \& White, S.D.M.
2002, ApJ, 576, 21
\bibitem{}
Vitvitska, M., Klypin, A.A., Kravtsov,  A.V., Bullock, J.S., Primack, 
J.R., Wechsler, R.H. 2002, ApJ, 581, 799
\bibitem{}
White, S.D.M. 1984, MNRAS, 286, 38
\bibitem{}
White, S.D.M. \& Rees, M.J. 1978, MNRAS, 183, 341
\bibitem{}
Zeldovich, Y.B. \& Novikov, I.D. 1983, in Relativistic Astrophysics, ed. G. Steigman
(Chicago: Univ. Chicago Press), p384
\end{chapthebibliography}
\end{document}